\documentclass[[aps,prd,a4paper,showpacs,twocolumn,groupedaddress]{revtex4-1}
\usepackage{amsmath,amssymb,bm}
\usepackage{gensymb}
\usepackage{nicefrac} 
\usepackage{epsfig}
\usepackage{graphicx}
\usepackage{slashed}
\usepackage{epsfig} \usepackage{graphicx} \usepackage{color}
\usepackage{mathrsfs} \usepackage{amssymb} \usepackage{amsmath} \usepackage{url}

\usepackage{xspace} 
\usepackage[bookmarks, breaklinks, colorlinks,urlcolor=black, citecolor=red, 
linkcolor=blue]{hyperref}

\def \c2b{\cos 2\beta}
\def \s2b{\sin 2\beta}

\def\sss{\scriptscriptstyle}

\def\CP {\ensuremath{C\!P}\xspace}

\def\beq{\begin{equation}}
\def\eeq{\end{equation}}
\def\bea{\begin{eqnarray}}
\def\eea{\end{eqnarray}}
\def\nn{\nonumber}

\def\sss{\scriptscriptstyle}
\def\roughly#1{\mathrel{\raise.3ex\hbox
{$#1$\kern-.75em\lower1ex\hbox{$\sim$}}}}

\def\Bpm{B_{+-}}
\def\Cpm{C_{+-}}
\def\Bzz{B_{00}}
\def\Czz{C_{00}}
\def\Bpz{B_{+0}}
\def\Cpz{C_{+0}}

\def\tda{2\Delta\alpha}

\def\Atpm{\tilde{A}^{+-}}
\def\Atzz{\tilde{A}^{00}}
\def\Atpz{\tilde{A}^{+0}}
\def\Atbpm{\tilde{\bar{A}}^{+-}}
\def\Atbzz{\tilde{\bar{A}}^{00}}
\def\Atbpz{\tilde{\bar{A}}^{+0}}

\def\PEWc{P_{EW}^C}

\allowdisplaybreaks

\begin{document}

\title{Constraining electroweak penguin graph contributions in measurements of the CKM phase alpha using $B\to\pi\pi$ and $B\to\rho\rho$ decays}

\author{Abinash Kumar Nayak}\email{abinashkn@imsc.res.in} 
\author{Rahul Sinha}\email{sinha@imsc.res.in}\affiliation{The Institute of 
Mathematical Sciences, Taramani, Chennai 600113, India \\ Homi Bhabha National 
Institute 
Training School Complex,  Anushakti Nagar, Mumbai 400085, India}
\author{Anirban Karan}\email{kanirban@iith.ac.in} 
\affiliation{Indian Institute of Technology, Hyderabad, Kandi, Sangareddy, 
Telangana 502285, India}
\author{Benjamin Grinstein}\email{bgrinstein@ucsd.edu}\affiliation{Department 
of 
Physics, UC San Diego, La Jolla, California 92093, USA}

\date{\today}

\begin{abstract}
The unitarity of the Cabibbo-Kobayashi-Maskawa (CKM) matrix has been well
established by both direct and indirect measurements without any evidence of
discrepancy. The CKM weak phase $\alpha$ is directly measured using an isospin
analysis in $B\rightarrow \pi\pi$ and $B\to \rho\rho$ assuming that  electroweak
penguin contributions are ignorable. However, electroweak penguins are sensitive
to NP, hence, it is important to experimentally estimate their effects. We
determine the size of both electroweak penguin and isospin amplitudes, directly
from  $B\rightarrow \pi\pi$ and $B\to \rho\rho$ experimental data, using in
addition the indirectly measured value of $\alpha$. We find that electroweak
penguin contribution are indeed small and agree with SM  expectations within
$1\sigma$. We also find that there is a mild enhancement of the $\Delta
I=\tfrac{1}{2}$ transition amplitude.
\end{abstract}


\maketitle

\section{Introduction}
\label{Intro}

The measurements of CKM phases (i.e., $\alpha, \beta$, $\gamma$) are very
crucial in understanding \CP violation, consequently a great deal of effort has
been put in over last few decades to measure them as accurately as possible. The
unitarity triangle obtained from these phase measurements is compared with other
indirect measurements~\cite{Charles:2004jd,Bona:2006ah}  to test for new physics
(NP) beyond the standard model (SM). At present no discrepancy has been observed
 between the direct and indirect measurements of the weak phases. The current
measurements are set to improve significantly given the large sample of data
expected at the LHCb and Belle\,II collaborations. 

While the measurements of weak phases have been the hallmark of Belle and $B\!A\!B\!A\!R$
collaboration, the methods that enabled the accurate measurements of weak phases
have marked an important era in the progress toward understanding CP violation. The
measurement of the weak phase $\alpha$ requires dealing with penguin
contributions that pollute this process, however, this issue is resolved by using an isospin
analysis~\cite{London:1989ph,Grinstein:1989df,Gronau:1989ia,Gronau:1990ka}. Indeed, the method of isospin analysis is used to measure $\alpha$ not
only using $B\to \pi\pi$ modes but also $B\to \rho\rho$ modes. The 
electroweak penguin could in principle also contribute to these modes and, again,
pollute the measurement of $\alpha$, but its contribution is expected to be
small within the SM. Since electroweak penguins are sensitive
to NP, it is important to experimentally estimate their effects. However, such 
an estimation is not possible using isospin
alone and requires one extra piece of information, as we will elaborate in 
detail. 

In this paper we have assumed SM and kept the electroweak penguin contributions. We then try to answer, how well the theory fits with the available experimental data. We make an assumption
that the indirect measurements of $\alpha$ ~\cite{Charles:2004jd,Bona:2006ah}
are indeed the correct value of $\alpha$. This indirect measurement of $\alpha$
readily provides the one extra piece of information. We estimate the size of the
electroweak penguin using data from both $B\to\pi\pi$ and $B\to\rho\rho$ modes.
We find that the electroweak penguin contributions are indeed small and in
$1\sigma$ agreement with theoretical expectations within the SM. Given the
current large errors in the measurements, there is neither any evidence of NP
nor any evidence of isospin violation. The measurement of time dependent
asymmetry in $B^0\to\rho^0\rho^0$ not only enables testing isospin but also
removes an ambiguity in the solution of the weak phase $\alpha$.

Our study also has particular relevance for $B\to\rho\rho$, since using the mode
involves several approximations. For instance, $\rho^0$ is a neutral vector meson and has
sizeable mixing with the photon resulting in long distance contributions that
can mimic contributions from the electroweak penguins. Also, a $I=1$ amplitude can, in principle, contribute resulting in corrections to the isospin analysis. Moreover, the
small contributions from transverse polarizations are ignored in the
experimental analysis. It is reassuring to find that $B\to \rho\rho$ also works well under these assumptions which gives us more confidence in the validity of these approximations. 

We take into account all possible penguin contributions into 
consideration and begin by describing the well known isospin in $B\to \pi\pi$ 
modes. The analysis for $B\to \rho\rho$ is similar. 
The $B\rightarrow \pi\pi$ amplitudes can in general be written as~\cite{Gronau:1995hn,Gronau:2001ff}
\begin{align}
	\label{eq:topology}
	\frac{1}{\sqrt{2}}A^{+-} &= (T+E) e^{i \gamma} + (P+\frac{2}{3} \PEWc) e^{-i 
	\beta}, \nonumber \\
	A^{00} &= (C-E) e^{i \gamma} +(P_{EW}+\frac{1}{3}\PEWc-P) e^{-i \beta}, 
	\nonumber \\
	A^{+0} &= (T+C) e^{i \gamma} + (P_{EW}+\PEWc)e^{-i \beta},  
\end{align}
where, $A^{+-}, A^{00}$, and $A^{+0}$ correspond to $B^0\rightarrow \pi^+ 
\pi^-$, $B^0\rightarrow \pi^0 \pi^0$, and $B^+\rightarrow \pi^+ \pi^0$, 
respectively. The complex topological amplitudes $T$, $C$ and $P$, $P_{EW}$, and 
$\PEWc$  indicate ``tree," ``color-suppressed-tree," ``penguin," 
``electroweak-penguin" and ``color-suppressed electroweak-penguin" amplitudes 
correspondingly and each of the amplitude includes the corresponding strong 
phases. There is also a smaller penguin annihilation amplitude which contributes to the 
$B^0$ decay modes and does not affect the isospin relation within the SM. It is customary to deal with redefined amplitudes where the amplitudes of the modes are rotated by $e^{-i\gamma}$ and those of the conjugate modes rotated by 
$e^{i\gamma}$, such that $\tilde{A}^{+-}=e^{-i\gamma} A^{+-}$ and 
$\tilde{\bar{A}}^{+-}=e^{i\gamma} \bar{A}^{+-}$, and the amplitudes 
$\tilde{A}^{00}$, $\tilde{A}^{+0}$ and $\tilde{\bar{A}}^{00}$, 
$\tilde{\bar{A}}^{+0}$ defined similarly. It is easy to see that no observables 
are altered by this redefinition.
We can cast the amplitudes in terms of $\alpha$ such 
that
\begin{align}
	\frac{1}{\sqrt{2}}\tilde{A}^{+-} &= (T+E) + X e^{i \alpha}, \nonumber \\
	\tilde{A}^{00} &= (C-E)+ Y e^{i \alpha},  \nonumber \\
	\tilde{A}^{+0} &= (T+C) + (X + Y )e^{i \alpha},
\end{align}
where, $X=(-P-\tfrac{2}{3} \PEWc)$ and $Y=(P-P_{EW}-\tfrac{1}{3}\PEWc)$. The 
conjugate amplitudes $\tilde{\bar{A}}^{+-}$, $\tilde{\bar{A}}^{00}$ and 
$\tilde{\bar{A}}^{+0}$ are obtained as usual from the amplitudes 
$\tilde{A}^{+-}$, $\tilde{A}^{00}$ and $\tilde{A}^{+0}$ by switching the sign 
of the weak phase $\alpha$.

An interesting point to note here is that $X+Y$ depends only on electroweak penguins $P_{EW}$ and the color suppressed counterpart $P^C_{EW}$~\cite{Gronau:1995hn}. Hence $X+Y$ serves as a measure of pure electroweak contributions in $B\rightarrow \pi\pi$. As evident from the above definitions, the amplitudes implicitly follow the isospin relations:
\begin{align}
	\label{T12}
	&\frac{1}{\sqrt{2}} \tilde{A}^{+-} + \tilde{A}^{00} = \tilde{A}^{+0} ,\nonumber 
	\\
	&\frac{1}{\sqrt{2}} \tilde{\bar{A}}^{+-} + \tilde{\bar{A}}^{00} = 
	\tilde{\bar{A}}^{+0} 
\end{align}
These two isospin relations in Eq.\eqref{T12} are inherently two triangle 
equations and the two triangles are described, up to a finite ambiguity, by the lengths of the 
sides and the relative angle between any related side of the two triangles. 
This requires ``seven'' measurements in total. The already  measured  branching 
fractions $B_{ij}$ as well as direct \CP asymmetries $C_{ij}$, defined as
\begin{figure}[t]
	\begin{center}
		\includegraphics[width=6cm]{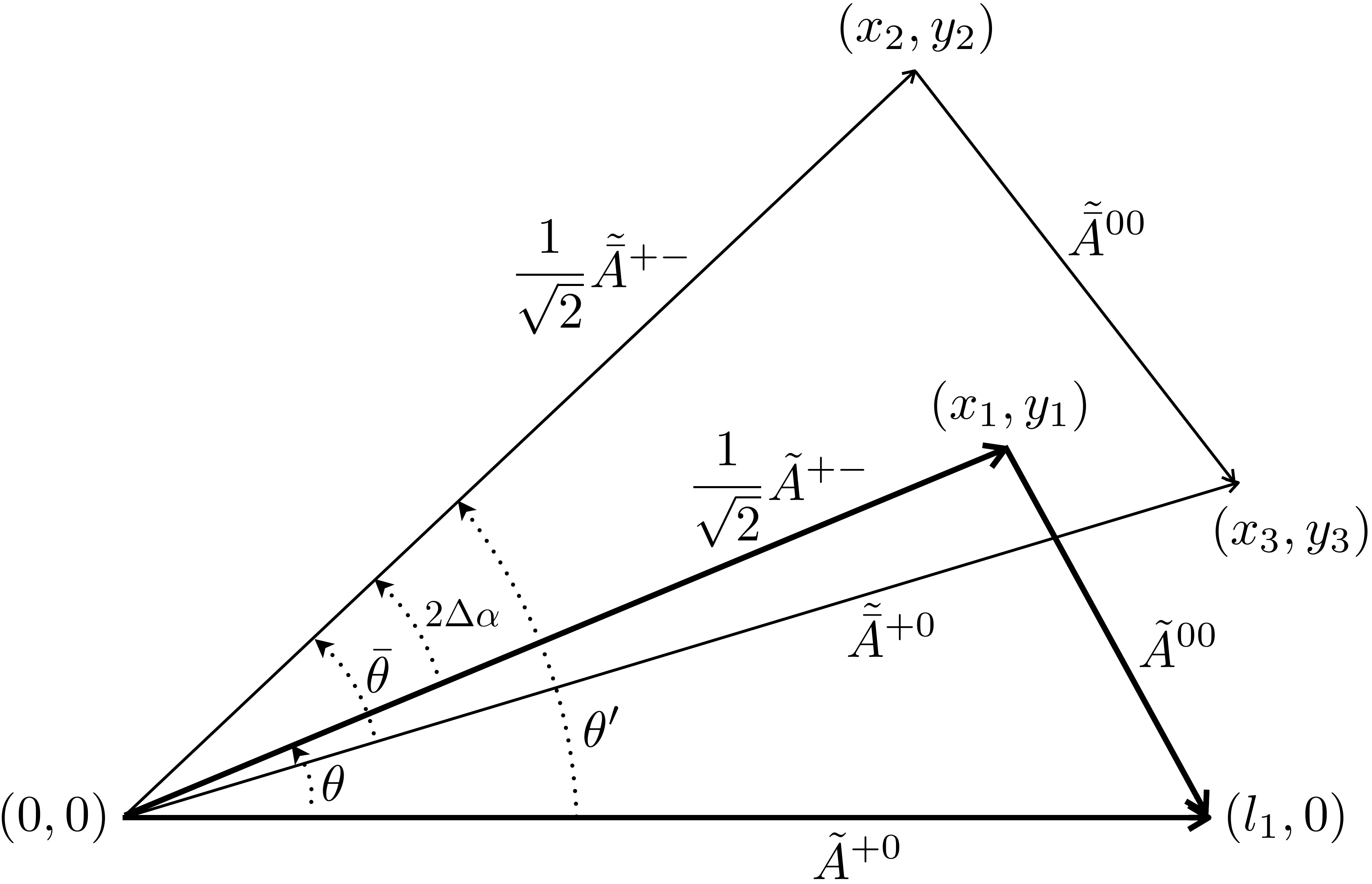}
		\caption{An illustration of the isospin triangles depicted in the complex 
		coordinate plane. 
		The figure defines the notation of coordinates and angles used to 
		obtain the 
		solutions of decay amplitudes including ambiguities. There is a 
		sixteen-fold 
		ambiguity in the solutions of coordinates as can be seen from 
		Eq.~\eqref{ASol}, 
		hence, there are sixteen distinct 
		orientations of the triangles drawn in this figure. However, only eight 
		solutions result in the correct value of $2\alpha^{\text{eff}} $.}
		\label{fig:two triangles}
	\end{center}
\end{figure}
\begin{equation}
	\label{eq:Br}
	B_{ij} = \frac{|\tilde{A}^{ij}|^2 + |\tilde{\bar{A}}^{ij}|^2}{2} , \qquad
	C_{ij} = \frac{|\tilde{A}^{ij}|^2 -|\tilde{\bar{A}}^{ij}|^2} 
	{|\tilde{A}^{ij}|^2 + |\tilde{\bar{A}}^{ij}|^2} ,
\end{equation}
provide complete information about each individual triangle. There is yet another measurement related to phase between $\tilde{A}^{ij}$ and $\tilde{\bar{A}}^{ij}$ obtained by the measurement of 
time-dependent \CP asymmetry in $B\rightarrow \pi^+\pi^-$, i.e., $S_{+-}$ which is defined as
\begin{align}
S_{+-} = \sqrt{1-C_{+-}^2} \sin(2\alpha^{\text{eff}}),
\end{align}
where, $2\alpha^{\text{eff}} = 2\alpha + 2\Delta\alpha$ or 
$\pi-2\alpha^{\text{eff}} =  2\alpha + 2\Delta\alpha$ and $2\Delta\alpha$ is
the phase between $\tilde{\bar{A}}^{+-}$ and $\tilde{A}^{+-}$. However, without
the measurement of $\alpha$, the measurement of $2\alpha^{\text{eff}}$, by
itself provides no information on $2\Delta\alpha$ and the isospin triangle 
cannot be drawn if there is an electroweak penguin contribution. Hence, we use the 
indirect measurement of $\alpha$ as an input to estimate $2\Delta\alpha$.   The 
two triangles then indicated by Eq.\eqref{T12} are presented in the 
coordinate framework diagrammatically in Fig.~\ref{fig:two triangles}. Conventionally, electroweak
penguins are ignored and the amplitudes $\tilde{A}^{+0}=\tilde{\bar{A}}^{+0}$,
which means that the corresponding sides of the two triangles overlap and the
two triangles with their relative orientation are fixed. This seventh
measurement, $\alpha^{\text{eff}}$ then directly enables the measurement of
$\alpha$ with ambiguities. In the presence of electroweak penguins
$P_{EW}+\PEWc\neq 0$, it is easily noted that there are seven independent
hadronic parameters and one cannot determine these seven parameters
as well as the weak phase $\alpha$ from only seven possible independent
measurements. We hence use the $\alpha$ obtained by indirect measurements and
translate the difference between ``direct'' and ``indirect'' measurements  
to a bound on $\Delta\alpha$ and the electroweak penguins.  

We can determine the magnitudes of the amplitudes $\tilde{A}^{ij}, 
\tilde{\bar{A}}^{ij}$, 
using Eq.~\eqref{eq:Br}, resulting in the two triangles (Fig.~\ref{fig:two 
triangles}), with the sides expressed in terms of coordinates as follows:
\begin{equation}\label{eq:coordinates}
	\begin{split}
		\tfrac{1}{2} |\Atpm|^2&= x_1^2 + y_1^2 = \tfrac{1}{2} \{ \Bpm(1+\Cpm)\} \\
		|\Atpz|^2 &= l_1^2 = \Bpz(1+\Cpz) \\
		|\Atzz|^2 &= (x_1-l_1)^2 + y_1^2 = \Bzz(1+\Czz) \\
		\tfrac{1}{2} |\Atbpm|^2 &= x_2^2 + y_2^2 = \tfrac{1}{2} \{ \Bpm(1-\Cpm)\} \\
		|\Atbpz|^2 &= x_3^2 + y_3^2 = \Bpz(1-\Cpz) \\
		|\Atbzz|^2 &= (x_3-x_2)^2 + (y_3-y_2)^2 = \Bzz(1-\Czz) 
	\end{split}
\end{equation}
The solutions for the 
coordinates in terms of experimental observables are given by,
\begin{equation}\label{ASol}
	\begin{gathered}
	l_1 =|\Atpz|\\ 
	x_1 = \tfrac{1}{\sqrt{2}} |\Atpm| \cos\theta \qquad 
	y_1 = \tfrac{1}{\sqrt{2}} |\Atpm| \sin\theta \\ 
	x_2 =  \tfrac{1}{\sqrt{2}} |\Atbpm| \cos\theta' \qquad 
	y_2 =  \tfrac{1}{\sqrt{2}} |\Atbpm| \sin\theta' \\
	x_3 =   |\Atbpz| \cos(\theta'-\bar{\theta}) \qquad 
	y_3 =  |\Atbpz| \sin(\theta'-\bar{\theta})
\end{gathered}
\end{equation}
\begin{table}[t]
	\centering
	\begin{tabular}{|c|c|c|}
		\hline 
		\hline
		& $B\to\pi\pi$ & $B\to\rho\rho$ \\ 
		\hline 
		$B_{+-}\times 10^{-5}$ & $0.512\pm 0.019$ & $2.77\pm 0.19$ \\ 
		\hline 
		$C_{+-}$ & $-0.31 \pm 0.05$ & $0.0\pm 0.09$ \\ 
		\hline 
		$S_{+-}$ & $-0.67\pm 0.06$ & $-0.14\pm 0.13$ \\ 
		\hline 
		corr$(C_{+-},S_{+-})$ & $0.21$ & $-0.02$ \\ 
		\hline 
		$B_{00}\times 10^{-5}$ & $0.159\pm 0.026$ & $0.096\pm 0.015$ \\ 
		\hline 
		$C_{00}$ & $-0.33\pm 0.22$ & $0.2\pm 0.9$ \\ 
		\hline 
		$S_{00}$ & - & $0.3\pm 0.7$ \\ 
		\hline 
		corr$(C_{00},S_{00})$ & - & $0.0$ \\ 
		\hline 
		$B_{+0}\times 10^{-5}$ & $0.55\pm 0.04$ & $2.4\pm 0.19$ \\ 
		\hline 
		$C_{+0}$ & $-0.03\pm 0.04$ & $0.05\pm 0.05$ \\ 
		\hline 
		$\alpha$ & \multicolumn{2}{c|}{$91.9\pm 3.0$} \\ 
		\hline
		\hline
	\end{tabular} 
	\caption{The table shows the used experimental values of the branching 
		fraction, direct $CP$ asymmetry and time-dependent $CP$ asymmetry of 
		$B\rightarrow \pi\pi$ and $B\rightarrow \rho\rho$ modes observed in 
		\cite{Charles:2004jd,Amhis:2016xyh,Amhis:2018udz,Tanabashi:2018oca}, 
		respectively. Note that in order to maintain consistency between the definitions of $C_{ij}$ in \cite{Charles:2004jd,Amhis:2016xyh,Amhis:2018udz,Tanabashi:2018oca} and Eq.~\eqref{eq:Br}, the signs of $C_{+0}$ in Table~\ref{tab:Expinput} has been reversed as compared to the values reported in \cite{Charles:2004jd,Amhis:2016xyh,Amhis:2018udz,Tanabashi:2018oca}.}
	\label{tab:Expinput}
\end{table}
\begin{figure}[!bh]
	\begin{center}
	\includegraphics*[width=1.68in]{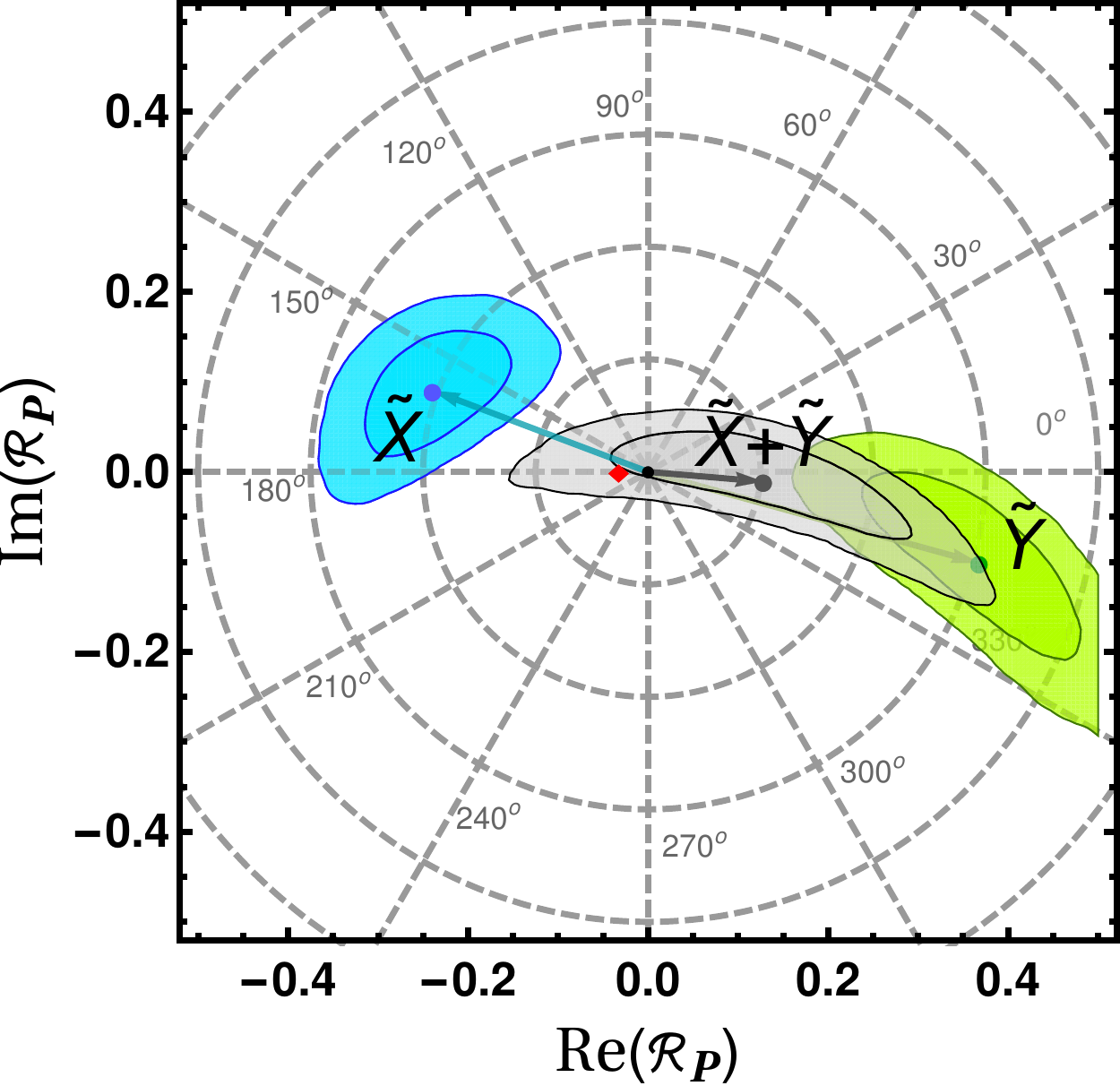}
	\includegraphics*[width=1.68in]{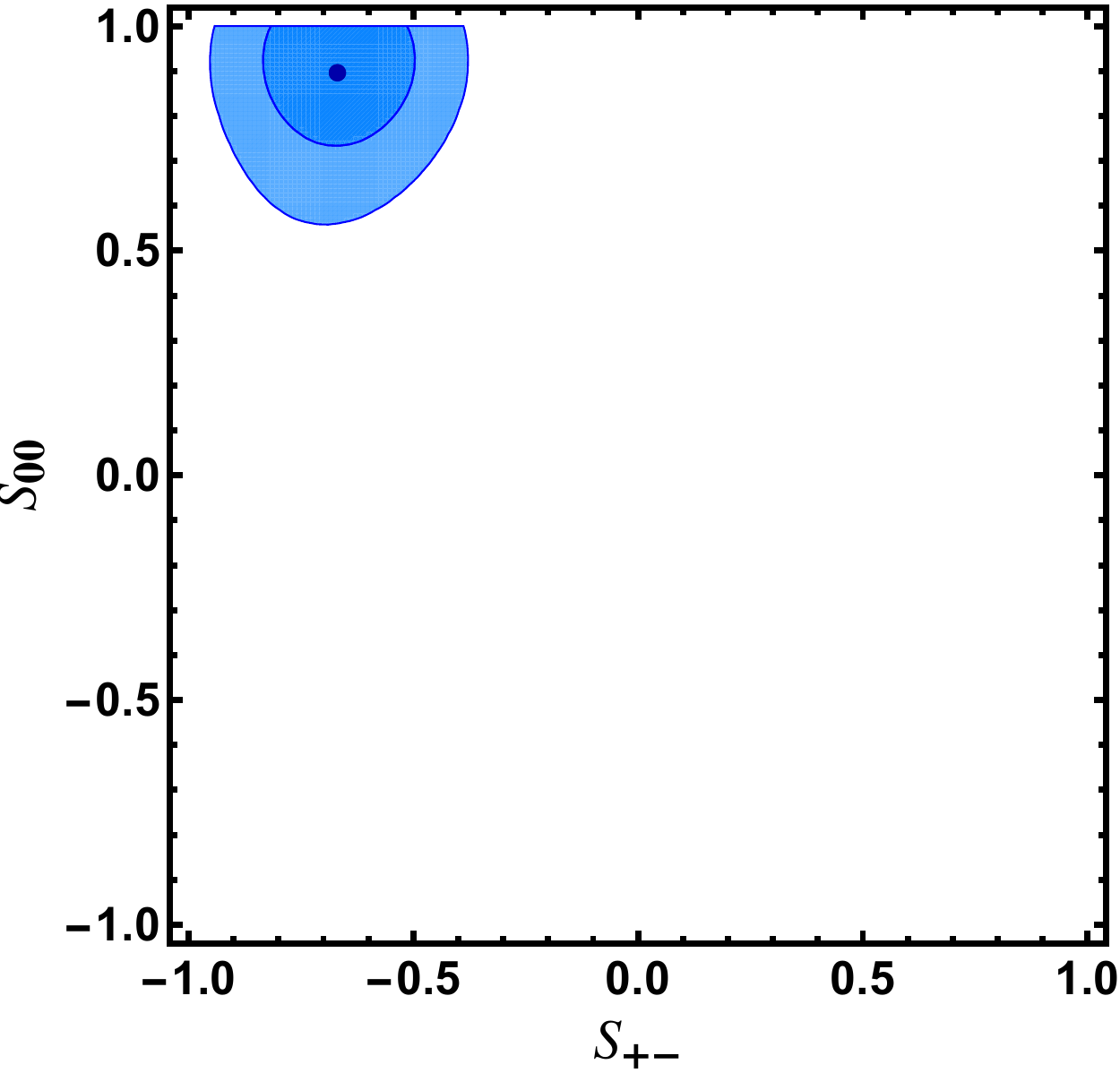} 
	\caption{The predicted 68.27$\%$ and 95.45$\%$ confidence levels for the
	topological amplitudes and $S_{00}$ versus $S_{+-}$ are illustrated for 
	$B\rightarrow \pi \pi$ modes. The light blue, light green, and light gray 
	contours correspond to the topological ratios $\tilde{X}$, $\tilde{Y}$ and 
	$\tilde{X}+\tilde{Y}$ respectively. The diamond ``${\color{red}\blacklozenge}$'' symbol at 
	$-0.0327$ represents the SM value of $z$. Out
	of the eight possible solutions, we have chosen to present the one where
	$\tilde{X}+\tilde{Y}$ is consistent with SM expectations.  The validity of this
	solution can easily be verified by a measurement of $S_{00}$, whose estimate is 
	shown in the figure on the right.}
	\label{fig:pitopology}
	\end{center}
\end{figure}
where $\cos\theta$ and $\cos\bar{\theta}$ are determined using cosine law in
terms of the amplitudes which are  expressed in terms of observables using
Eq.~\eqref{eq:coordinates}. $\sin\theta$ and $\sin\bar{\theta}$ are then each
obtained up to a two fold ambiguity. The other unknown $\theta' = \theta + \tda$,
where the phase $\tda$ itself has a two-fold ambiguity and is given by $\tda =
2\alpha^{\text{eff}}-2\alpha$ or $\tda =\pi- 2\alpha^{\text{eff}}-2\alpha$. It
is easily seen from Eq.~\eqref{ASol} that there is an sixteen-fold ambiguity in
the solutions of the coordinates. However, we find that only eight solutions
result in the correct value of $2\alpha^{\text{eff}}$, resulting in an
eight-fold ambiguity in the solution to the amplitudes.  It is well known that
$\alpha$ can be measured with up to eight-fold ambiguity using the conventional
technique. Hence, a eight-fold ambiguity in the determination of
decay amplitudes $\tilde{A}^{ij}$, $\tilde{\bar{A}}^{ij}$ is consistent with
expectation.

\begin{figure}[tb]
	\begin{center}
	\includegraphics*[width=1.67in]{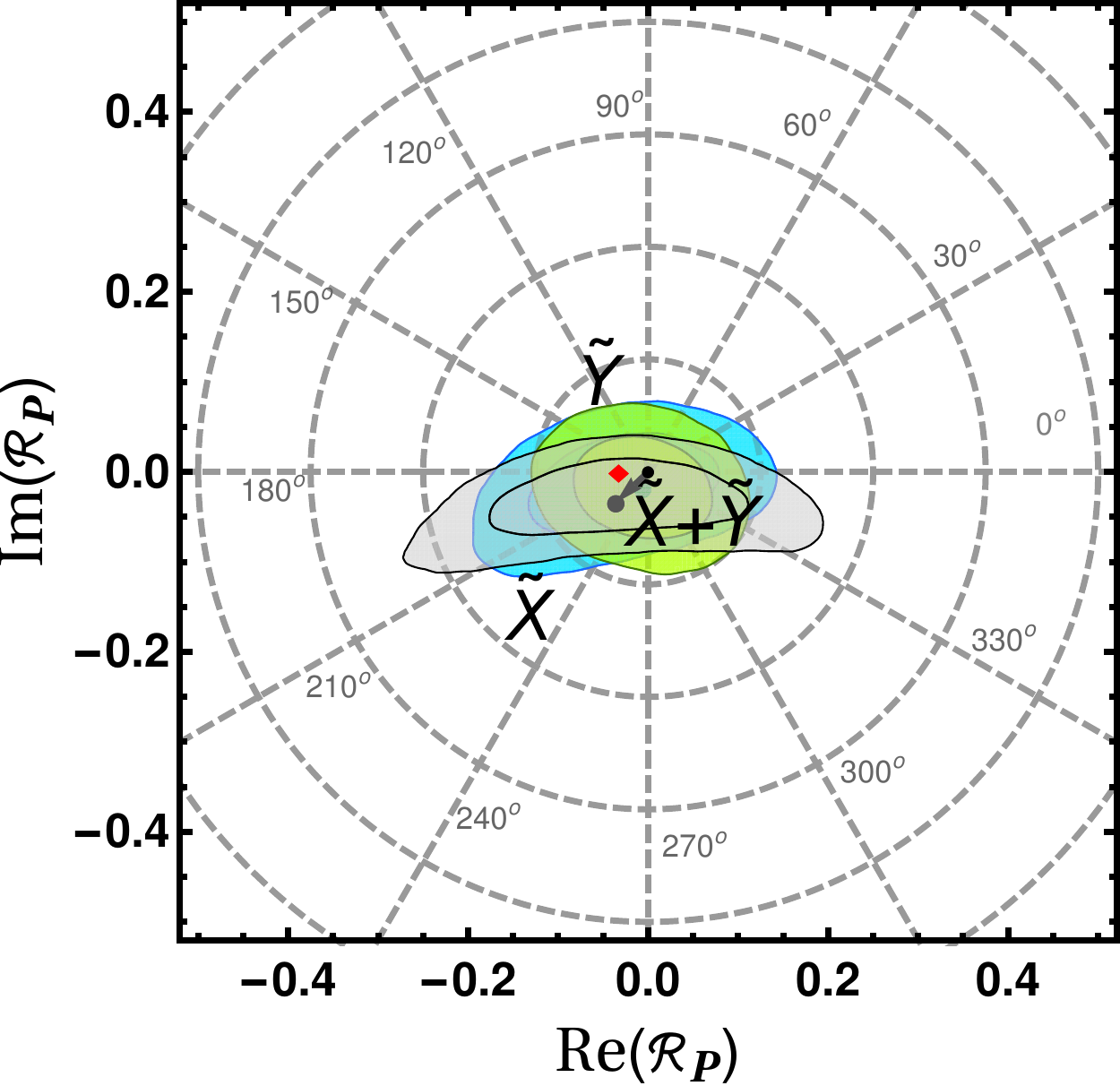}
	\includegraphics*[width=1.68in]{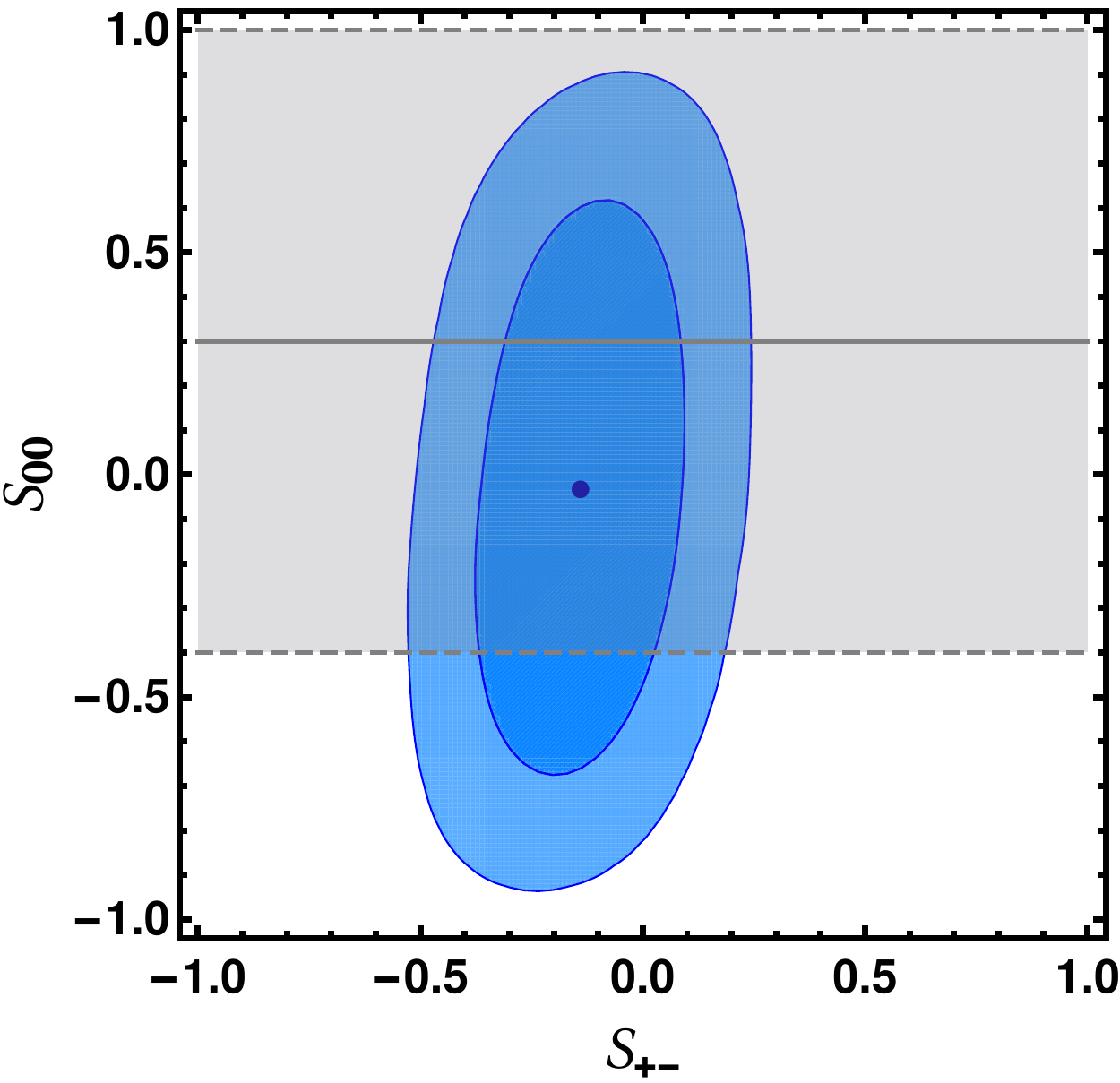}
	\includegraphics*[width=1.67in]{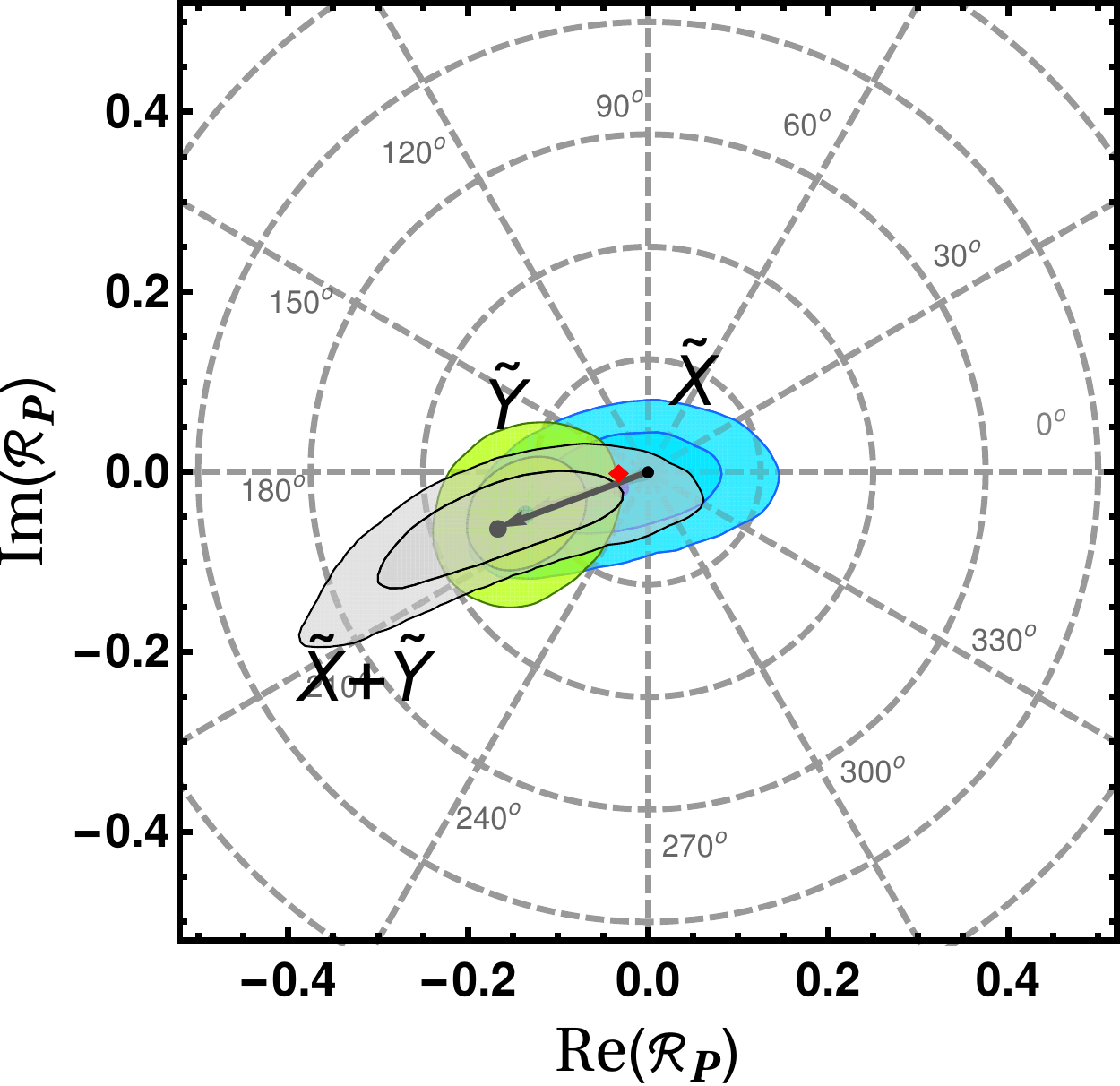}
	\includegraphics*[width=1.68in]{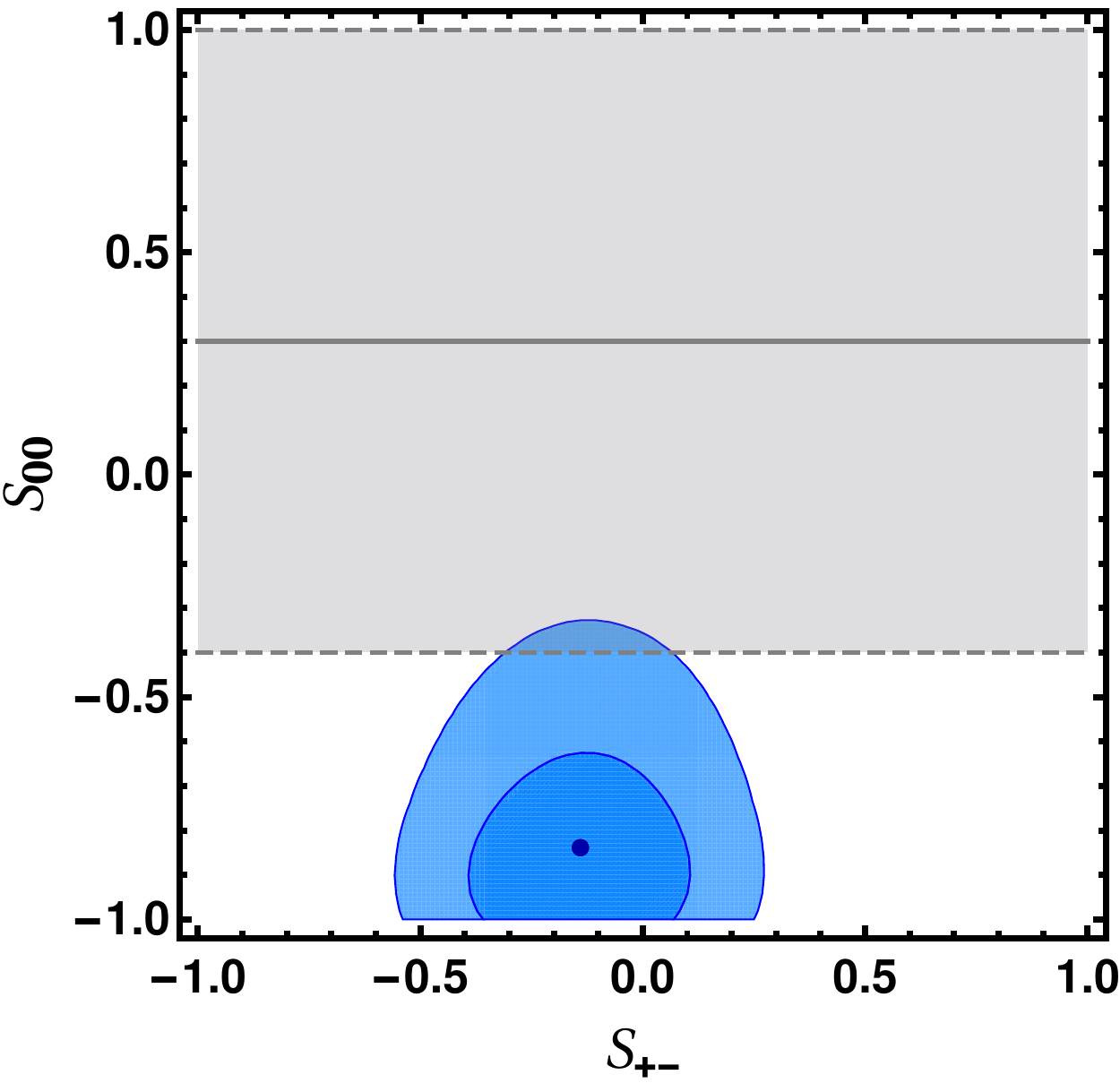}
	\includegraphics*[width=1.67in]{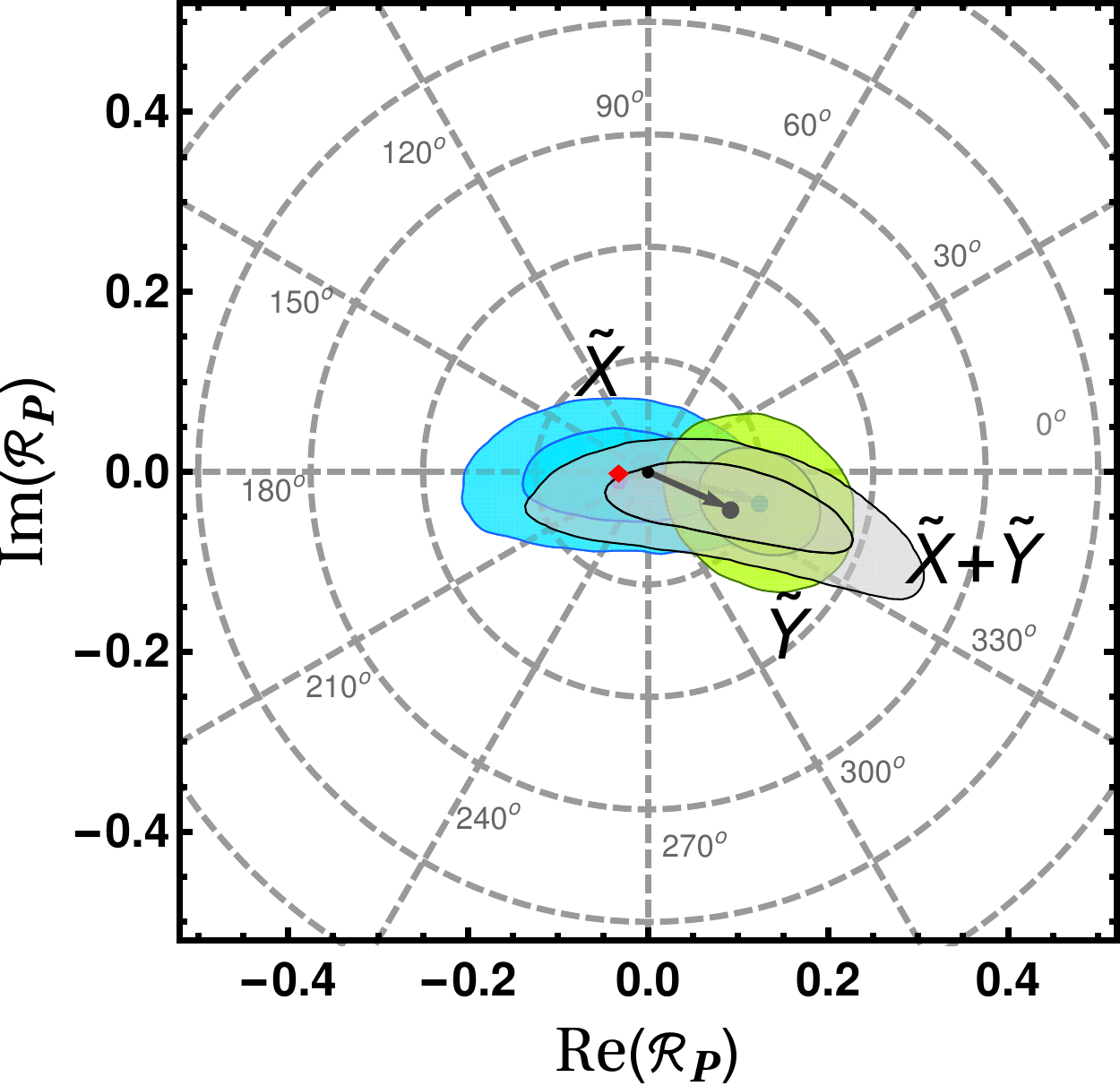}
	\includegraphics*[width=1.68in]{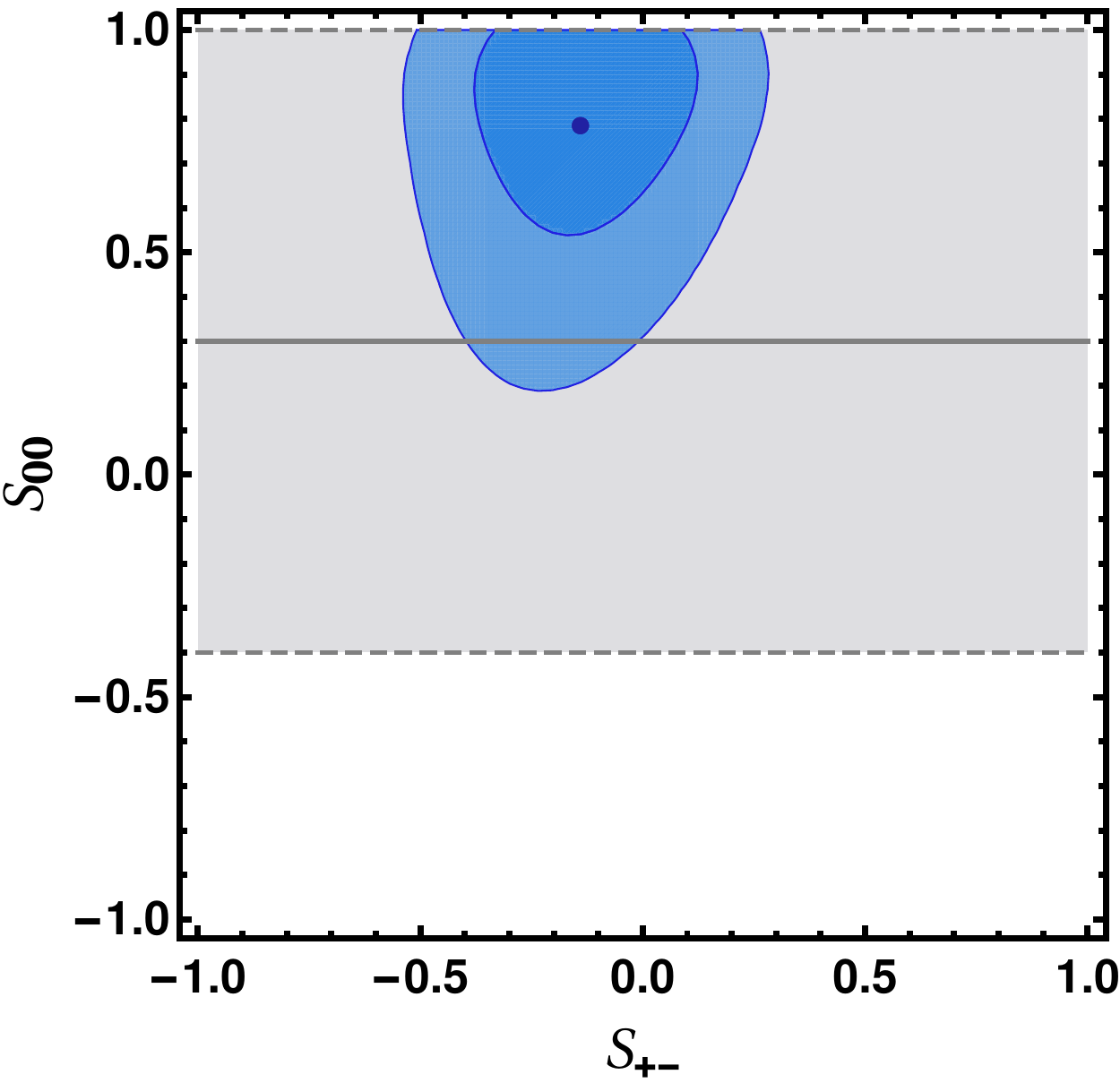}
	\includegraphics*[width=1.67in]{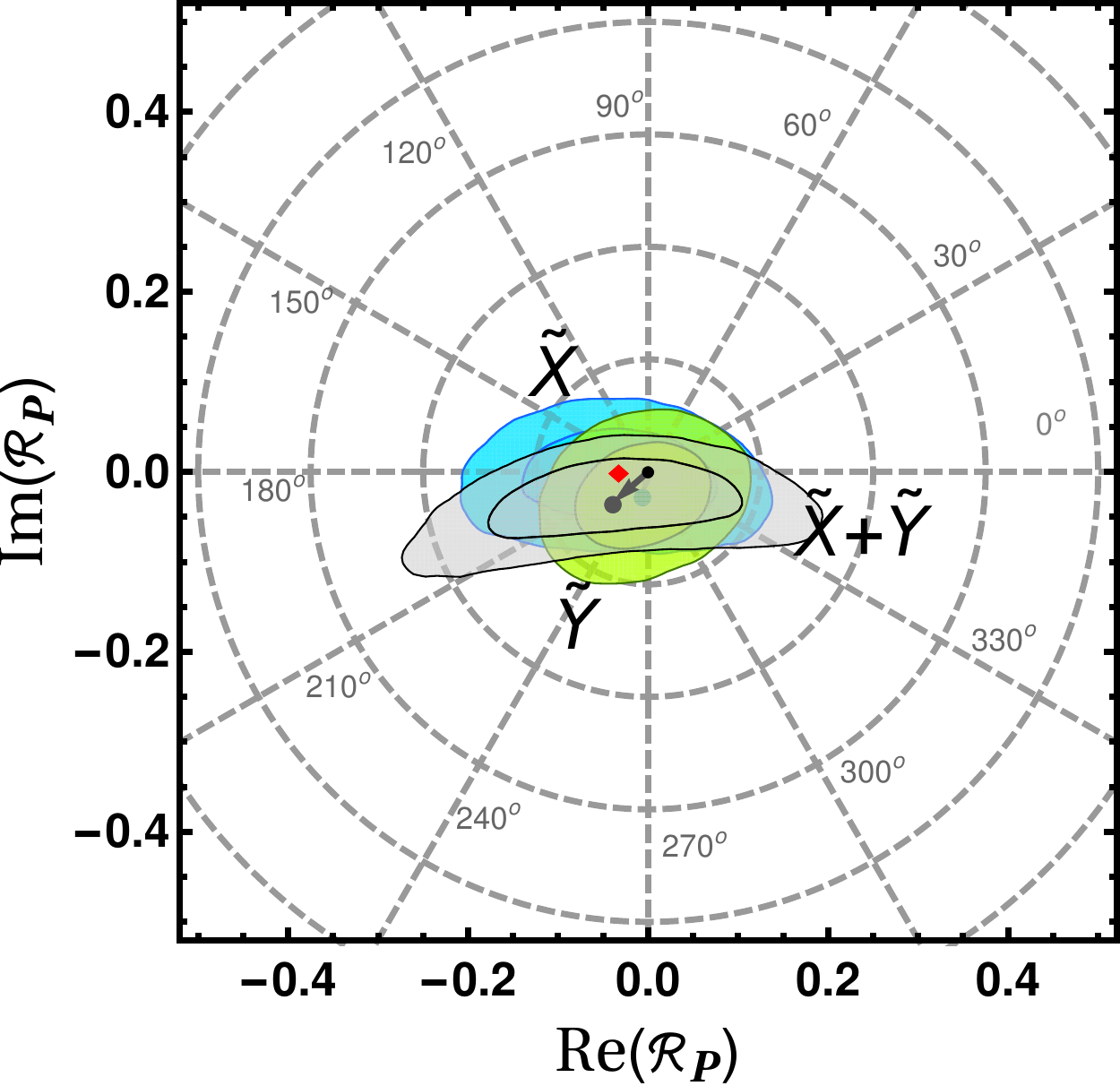}
	\includegraphics*[width=1.68in]{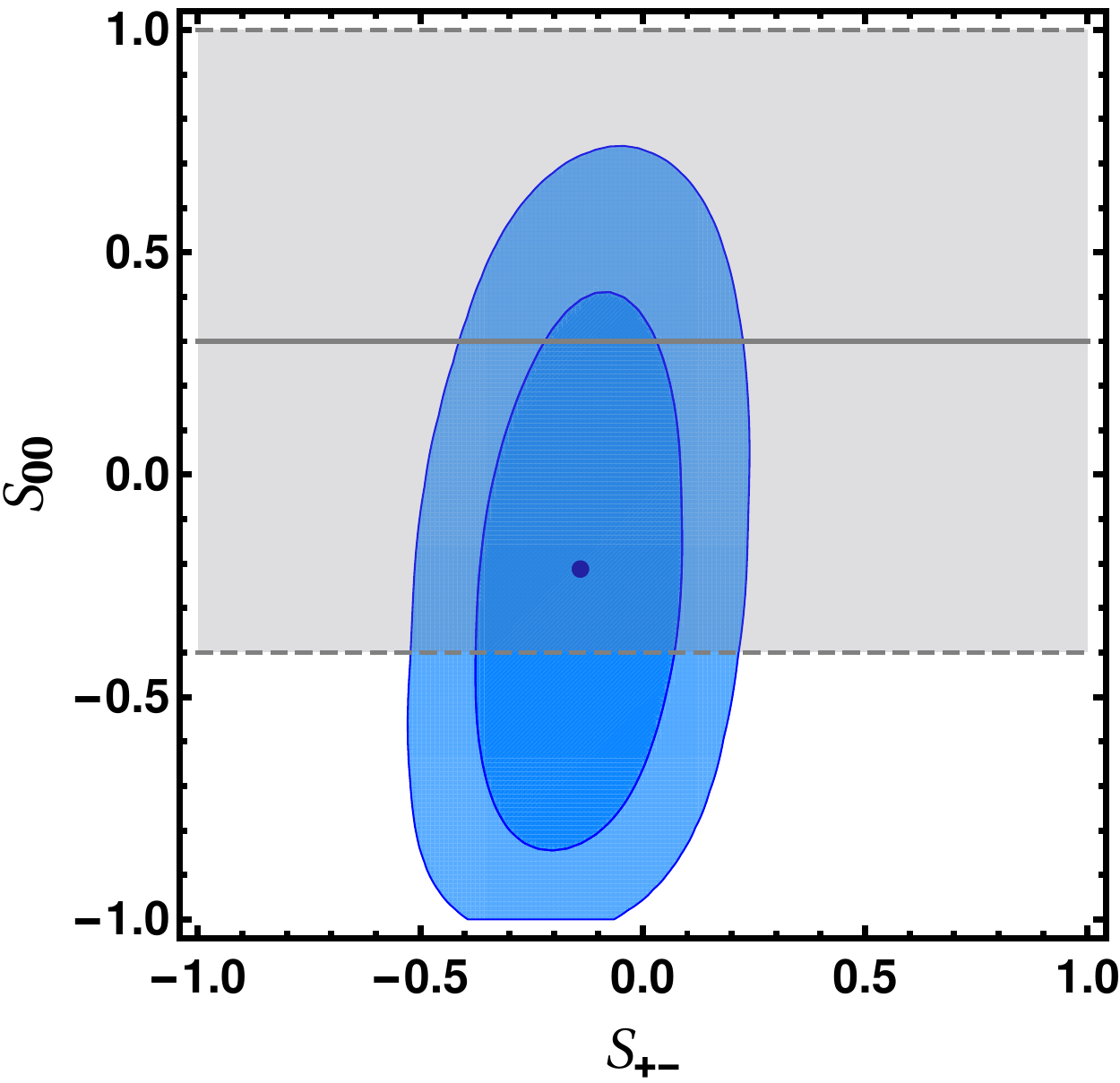}
	\caption{The topological amplitudes and $S_{00}$ versus $S_{+-}$ are illustrated
	for $B\rightarrow \rho \rho$ modes. The details of the contours are the same an
	in Fig.~\ref{fig:pitopology}.  Notice that $\tilde{X}+\tilde{Y}$, is consistent with SM 
	expectations for four of the solutions.  With an accurate
	measurement of $S_{00}$, whose estimate is shown in the figure on the right, the
	correct ambiguity can be identified. $1\sigma$ bands for the measurement of
	$S_{00}$ are superimposed along with our estimates for ready reference.}
	\label{fig:rhotopology}
	\end{center}
\end{figure}

In Table~\ref{tab:Expinput} we have summarized the experimental inputs used in
our analysis to  generate the data sets as normal distributions around the
observed central values with errors and available correlations. Furthermore, we
ensure that the simulated data sets are in compliance with Eq.\eqref{T12} by
imposing triangle inequalities for the respective triangles. In addition, $-1
\leq \{C_{ij},S_{ij}\} \leq 1$ has been implemented to allow only physically
allowed values. For each choice of data set satisfying the constraints, we
obtain eight possible equivalent solutions for the amplitudes. We find that for
$B\to \pi\pi$ that the triangles obtained by simulated amplitudes close in only
about half of the cases. Whereas, for the $B\to \rho\rho$ mode the valid cases
reduce to only a few percent. The closure of the isospin triangles is ensured by
the isospin bounds~\cite{Gronau:2001ff,Falk:2003uq} on $B_{00}$ and the 
observed values of
$B_{00}$ are very small and barely satisfy the isospin bounds for both $B\to
\pi\pi$ and $B\to \rho\rho$ modes.

Having determined the complex decay amplitudes, the topological
amplitudes $T+E$, $C-E$, $X$, $Y$ and the observable $S_{00}$ can all be
determined for each data set. Our 
interest is in estimating the size of the electroweak penguin. We hence 
determine, the ratios of the penguin contributions compared to the tree 
contributions generically denoted by ${\cal 
R}_P=\{\tilde{X},\tilde{Y},\tilde{X}+\tilde{Y}\}$ as 
\begin{equation}\label{eq:RP}
	\begin{gathered}
	\tilde{X}=\displaystyle\frac{X}{|T+C|}\qquad
	\tilde{Y}=\displaystyle\frac{Y}{|T+C|}\\[2ex]
	\tilde{X}+\tilde{Y}=\displaystyle\frac{X+Y}{|T+C|}\equiv z 
	e^{i\delta_{\sss{T\!C}}},
	\end{gathered}
\end{equation}
where $z$ is defined in Eq.~\eqref{eq:z} and $\delta_{\sss{T\!C}}$ is the 
strong phase of $T+C$.

The parameter $z$ has been theoretically estimated earlier. It is well known 
that only the $\Delta I=\tfrac{3}{2}$ part of the Hamiltonian 
contributes to the decay $B^\pm\to \pi^\pm \pi^0$, and the tree and electroweak 
part of the $\Delta I=\tfrac{3}{2}$ Hamiltonian are 
related~\cite{Gronau:1998fn} 
assuming only that $C_7$ and $C_8$ can be neglected, as follows:
\begin{equation}
	\label{eq:Hamiltonian}
	{\cal H}^{\sss EW}_{\sss \Delta 
	I=\tfrac{3}{2}}=-\frac{3}{2}\frac{V_{tb}V_{td}}{V_{ub}V_{ud}}\frac{C_9+C_{10}}{C_1+C_2}
 	{\cal H}^{\sss\text {tree}}_{\sss\Delta I=\tfrac{3}{2}}
\end{equation}
The equality $\tilde{\bar{A}}^{+0}=\tilde{A}^{+0}$ is broken 
by electroweak penguins and these amplitudes are expressed as
\begin{align}
	\tilde{A}^{+0} &= (T+C) + z e^{i \alpha}(T+C),\\ 
	\tilde{\bar{A}}^{+0} &= (T+C) + z e^{-i \alpha} (T+C), 
\end{align}
where,
\begin{equation}
	\label{eq:z}
	z=-\frac{3}{2}\Bigg|\frac{V_{tb}V_{td}}{V_{ub}V_{ud}}\Bigg|
		\frac{C_9+C_{10}}{C_1+C_2}  \approx -0.013 
		\Bigg|\frac{V_{tb}V_{td}}{V_{ub}V_{ud}}\Bigg|.
\end{equation}
The value of ratio of CKM elements $(V_{tb}V_{td})/({V_{ub}V_{ud}})$ is 
obtained from Ref.\cite{Charles:2004jd}.

The  $68.27\%$ and $95.45\%$ confidence levels for ${\cal R}_P$ obtained from
the probability distribution functions are shown in Fig.~\ref{fig:pitopology} and
Fig.~\ref{fig:rhotopology} for $B\rightarrow \pi\pi$ and $B\to\rho\rho$ decay modes
respectively. Also plotted are the corresponding estimates for the observable
$S_{00}$, derived from the amplitudes. It can be seen that if $S_{00}$ is
measured some of the solutions can be eliminated. In Fig.~\ref{fig:pitopology}, we have
chosen to present only one out of the eight possible solutions where
$\tilde{X}+\tilde{Y}$ is in agreement with the SM estimate within one standard
deviation. Measurements of the associated time-dependent \CP asymmetry $S_{00}$ can reduce or even eliminate the ambiguity.

\begin{figure}[!ht]
	\begin{center}
		\includegraphics[width=6cm]{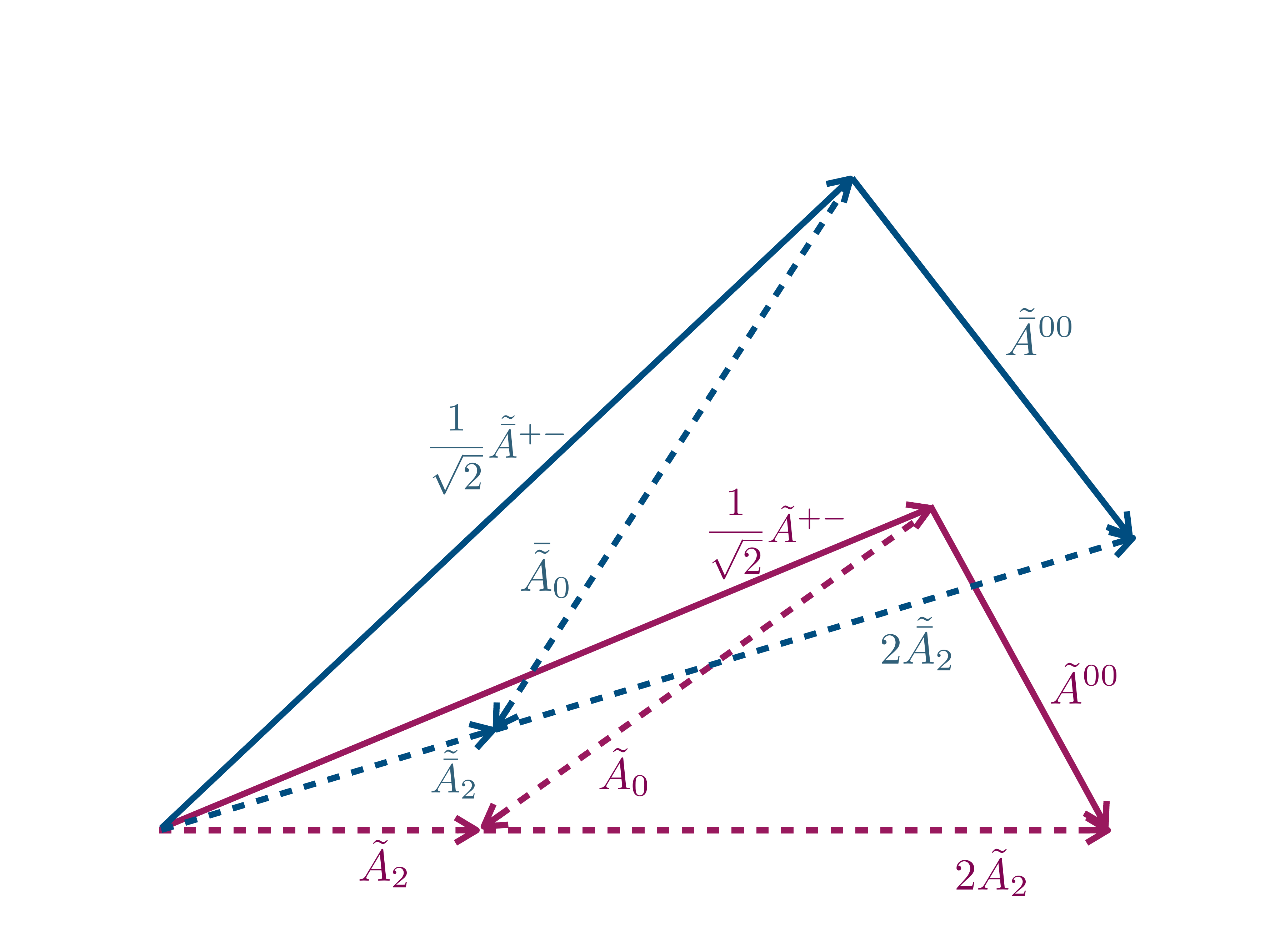}
		\caption{An illustration of the isospin triangles depicted in isospin space. 
		the isospin amplitudes $\tilde{A}_0$, $\tilde{A}_2$, 
		$\tilde{\bar{A}}_0$ and  $\tilde{\bar{A}}_2$ defined in Eq.~\eqref{eq:isospin} 
		are illustrated here.}
		\label{fig:isospin}	
	\end{center}
\end{figure}

The rotated amplitudes $\tilde{A}^{ij}$ for the decay $B\to \pi\pi$ can also be 
decomposed \cite{Gronau:1990ka} in terms of $I=0$ and $I=2$ isospin amplitudes 
as follows:
\begin{align}
	\label{eq:isospin}
	\frac{1}{\sqrt{2}}\tilde{A}^{+-} &=\tilde{A}_2-\tilde{A}_0\nn\\
	\tilde{A}^{00}&=2 \tilde{A}_2+\tilde{A}_0\nn\\
	\tilde{A}^{+0}&=3 \tilde{A}_2,
\end{align}
with analogous expressions for the three conjugate mode amplitudes 
$\tilde{\bar{A}}^{ij}$. A graphical representation of Eq.~\eqref{eq:isospin} is 
shown in Fig.~\ref{fig:isospin}. 
\begin{figure}[h]
	\begin{center}
		\includegraphics*[width=1.55in]{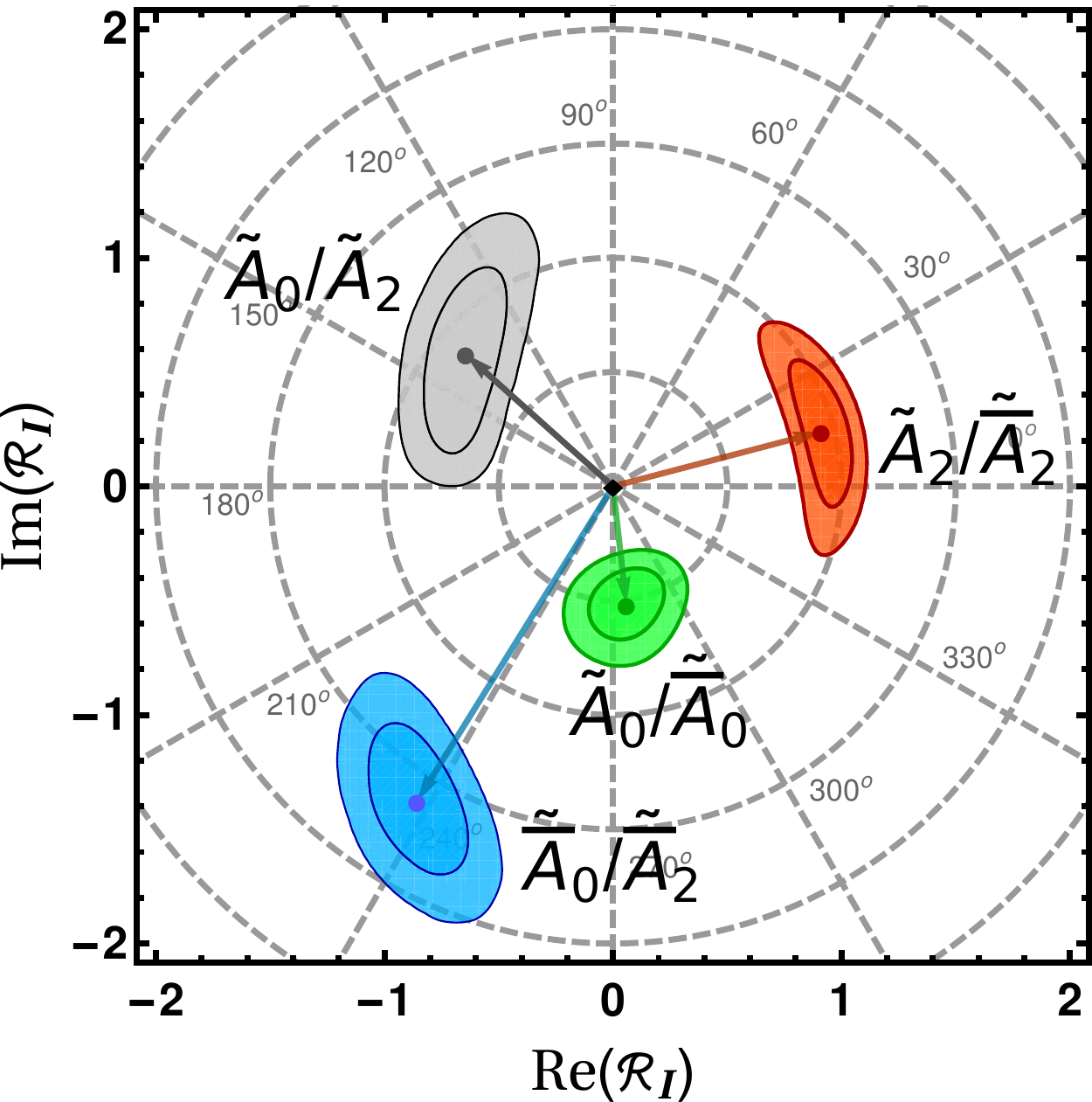}
		\caption{The predicted $68.27\%$ and $95.45\%$ confidence levels of isospin
		ratios for  $B\rightarrow \pi \pi$ modes. The gray, blue, green and orange
		contours correspond to $A_0/A_2$, $\bar{A}_0/\bar{A}_2$, $A_0/\bar{A}_0$, and
		$A_2/\bar{A}_2$ isospin amplitude ratios, respectively.The solution presented
		corresponds to ambiguity presented in Fig.~\ref{fig:pitopology}} 
		\label{fig:piisospin}
	\end{center}
\end{figure}
\begin{figure}[!t]
	\begin{center}
		\includegraphics*[width=1.4in]{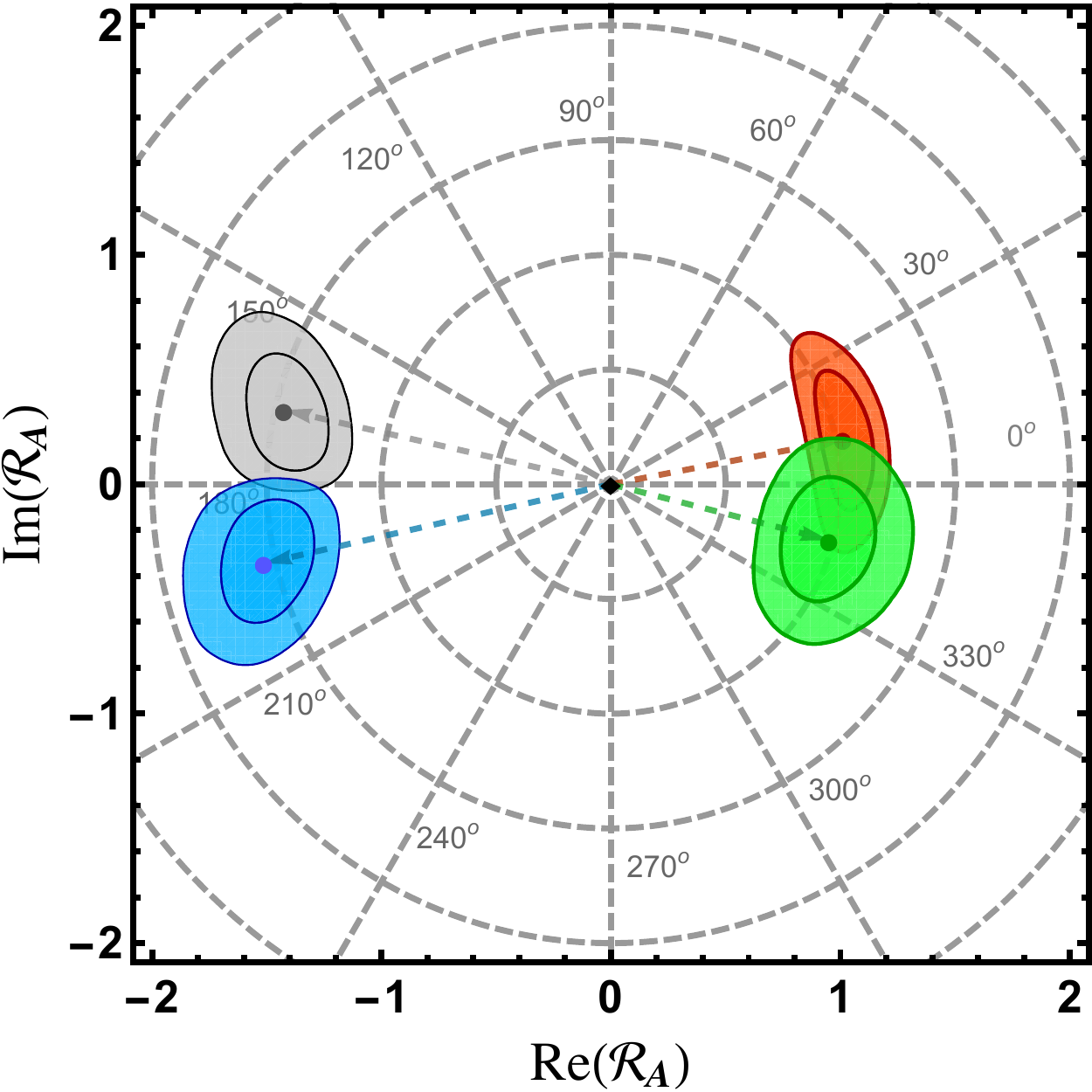}
		\includegraphics*[width=1.4in]{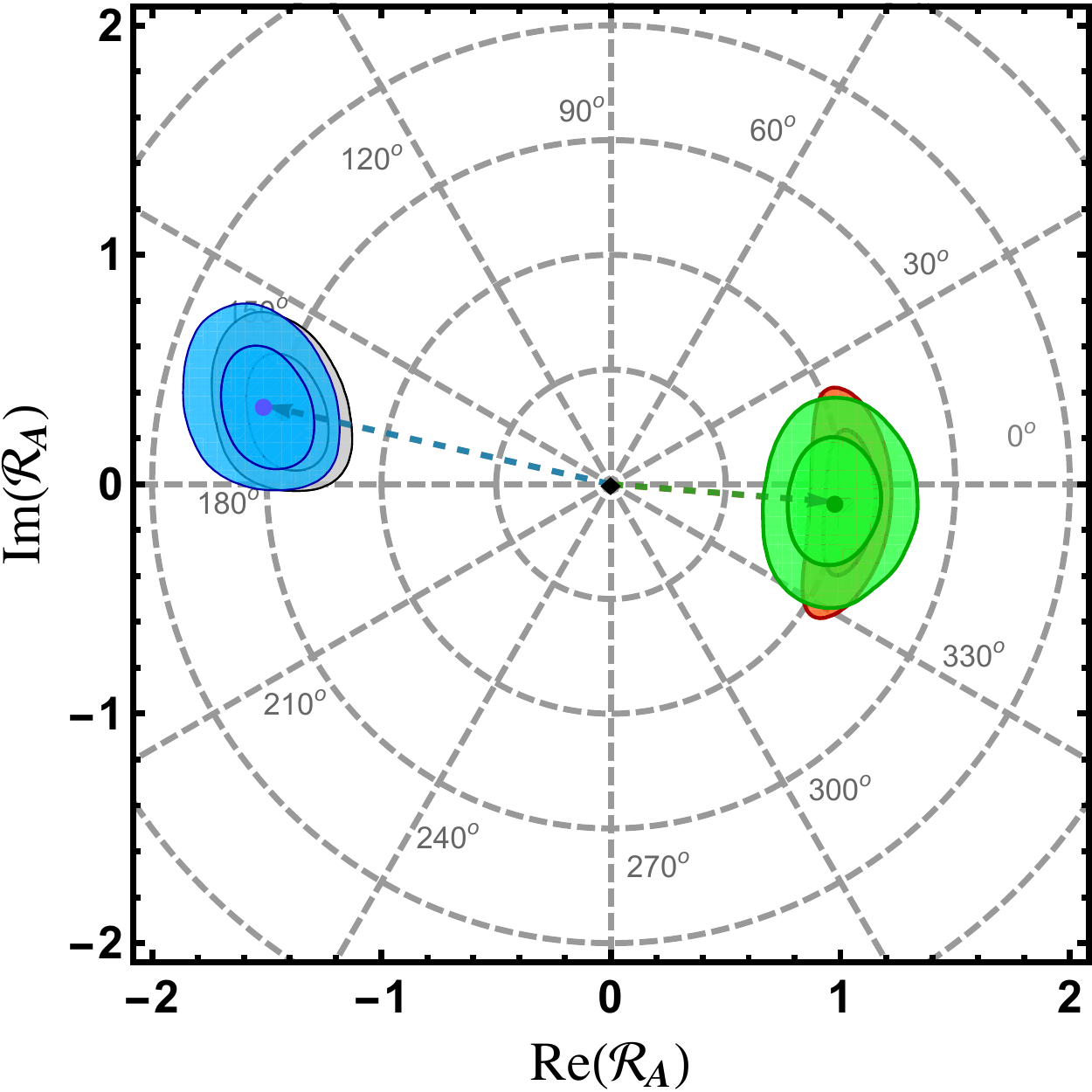}
		\caption{The predicted isospin amplitudes ratios for $B\rightarrow \rho \rho$ 
		modes. See Fig.~\ref{fig:piisospin} for details. Two other solutions correspond 
		to reflections around the horizontal axis.} 
		\label{fig:rhoisospin}
	\end{center}
\end{figure}
The measurements of the seven observables 
enable the 
complete 
determination of the four isospin amplitudes $\tilde{A}_0$, $\tilde{A}_2$, 
$\tilde{\bar{A}}_0$ and  $\tilde{\bar{A}}_2$. 
The isospin amplitudes $\tilde{A}_0$ and $\tilde{A}_2$ are easily written in  
terms of the topological amplitudes as follows:
\begin{align}\label{eq:iso-top}
	\tilde{A}_0 &=\frac{C-2T-3E}{3}+\frac{Y-2X}{3}e^{i\alpha}\\
	\tilde{A}_2 &=\frac{C+T}{3}+\frac{X+Y}{3}e^{i\alpha}
\end{align} 
We have studied the ratios of the isospin amplitudes generically denoted by 
 $${\cal R}_I=\{\tilde{A}_0/\tilde{A}_2, \tilde{\bar{A}}_0/\tilde{\bar{A}}_2, 
 \tilde{A}_0/\tilde{\bar{A}}_0, \tilde{A}_2/\tilde{\bar{A}}_2 \}$$
Note that 
$\tilde{A}_0/\tilde{A}_2= A_0/A_2$ and $\tilde{\bar{A}}_0/\tilde{\bar{A}}_2=  
\bar{A}_0/\bar{A}_2$.

We find that for  $B\to \pi\pi$ the hierarchy of isospin amplitudes is   
$|A_2|\approx |\bar{A}_2| \lesssim |A_0|<|\bar{A}_0|$ whereas for $B\to \rho\rho$  it 
follows that
$|A_2|\approx |\bar{A}_2| < |A_0| \approx |\bar{A}_0|$. These observations can be 
easily verified 
from Fig.~\ref{fig:piisospin} and \ref{fig:rhoisospin}. The right side plot of Fig.~\ref{fig:rhoisospin} 
deserves special consideration. It is easy to see that $A_0/A_2$ 
and $\bar{A}_0/\bar{A}_2$ can be written in 
terms of the topological amplitudes and has the form 
\begin{align}\label{eq:A0byA2exp}
	\frac{A_0}{A_2} &= xe^{i\delta_x}+iy e^{i\delta_y}\\
	\frac{\bar{A}_0}{\bar{A}_2} &= xe^{i\delta_x}-iy e^{i\delta_y}
\end{align}
where $x$, $y$, $\delta_x$ and $\delta_y$ are complicated function of 
topological amplitudes and $\alpha$. Hence, the overlapping plots seen the 
right side figure in Fig.~\ref{fig:rhoisospin} happen if
\begin{equation}
	\frac{A_0}{A_2}\approx \frac{\bar{A}_0}{\bar{A}_2}\implies y=0\implies
	\frac{C-E}{T+E}\approx \frac{Y}{X}.
\end{equation}

To conclude we have shown that assuming the value of $\alpha$ obtained from
indirect measurements and available experimental data for $B\to\pi\pi$ and
$B\to\rho\rho$ observables, all topological and isospin amplitudes can be
extracted. These solutions come with an eightfold ambiguity, and only one solution yields small values of electroweak penguins, consistent with SM expectations. Measurements of the associated time-dependent \CP asymmetry $S_{00}$ can reduce or even eliminate the ambiguity. The interesting
conclusion drawn is that the size of that electroweak penguin contributions are
consistent with theoretical expectations given the current experimental
uncertainties.  Improved accuracy in the measurements of observables for these
modes and of the indirect measurement of $\alpha$ will help in understanding the
electroweak penguin contributions to hadronic modes. We also find a hierarchy
among the isospin amplitudes with  mild enhancement of the $\Delta
I=\tfrac{1}{2}$ transition amplitude.

{\em Acknowledgement}: A.K.N. and R.S. would like to thank Sunando Kumar Patra 
for useful discussions. B.G. thanks Institute of Mathematical Sciences for 
hospitality where part of the work was done. B.G. was supported in part
by the US Department of Energy grant No. de-sc0009919. A.K. thanks SERB India, grant no: SERB/PHY/F181/2018-19/G210, for support.
R.S. thanks Perimeter Institute for Theoretical Physics for hospitality where 
part of this work was done. Research at Perimeter Institute is supported by the 
Government of Canada through the Department of Innovation, Science and Economic 
Development and by the Province of Ontario through the Ministry of Research, 
Innovation and Science.

\end{document}